# A pulse oximeter based on Time-of-Flight histograms

Yuanyuan Hua*[a], Konstantinos Bantounos [a], Aravind Venugopalan Nair Jalajakumari [a], Alex Turpin[b], Ian Underwood [a], Danial Chitnis [a]

[a] School of Engineering, Institute for Integrated Micro and Nano Systems, University of Edinburgh, Edinburgh, UK, EH9 3FF;
[b] School of Computing Science, University of Glasgow, Glasgow, UK, G12 8QQ;

**ABSTRACT**

A pulse oximeter is an optical device that monitors tissue oxygenation levels. Traditionally, these devices estimate the oxygenation level by measuring the intensity of the transmitted light through the tissue and are embedded into everyday devices such as smartphones and smartwatches. However, these sensors require prior information and are susceptible to unwanted changes in the intensity, including ambient light, skin tone, and motion artefacts. Previous experiments have shown the potential of Time-of-Flight (ToF) techniques in measurements of tissue hemodynamics. Our proposed technology uses histograms of photon flight paths within the tissue to obtain tissue oxygenation, regardless of the changes in the intensity of the source. Our device is based on a 45ps time-to-digital converter (TDC) which is implemented in a Xilinx Zynq UltraScale+ field programmable gate array (FPGA), a CMOS Single Photon Avalanche Diode (SPAD) detector, and a low-cost compact laser source. All these components including the SPAD detector are manufactured using the latest commercially available technology, which leads to increased linearity, accuracy, and stability for ToF measurements. This proof-of-concept system is approximately 10cm×8cm×5cm in size, with a high potential for shrinkage through further system development and component integration. We demonstrate preliminary results of ToF pulse measurements and report the engineering details, trade-offs, and challenges of this design. We discuss the potential for mass adoption of ToF based pulse oximeters in everyday devices such as smartphones and wearables.

## 1. INTRODUCTION

A pulse oximeter is a vital device which monitors the oxygen saturation level of hemoglobin in arterial blood. It is based on the wavelength dependence photon absorption of oxyhemoglobin (O2Hb) and deoxyhemoglobin (HHb). Typically, the absorption of the two wavelengths is used to calculate the ratio which determines oxygen levels. These wavelengths which are generally 660 nm (red) and 940 nm (near infrared) are used to illuminate the skin and photodiodes are applied to collect the reflected or transmitted photons [1]. However, these sensors measure only the intensity of the transmitted light and are susceptible to unwanted changes in the intensity, including motion artefacts [2]. Additionally, Research shows that skin tone may affect the accuracy of hemodynamic measurements in intensity-based pulse oximeters, which means prior information is needed to correct the measured data [3]. Compared to a conventional photodiode, which is traditionally used in the pulse oximeters to measure only intensity information, a Single Photon Avalanche Diode (SPAD) enables Time-of-Flight (ToF) measurements which provides both intensity and temporal information of photon flight time. Previous experiments have shown the potential of ToF techniques in measurement of tissue hemodynamics [4][5][6].

In this paper, the ToF technology we propose uses real-time histograms of photon flight paths within the tissue to obtain the preliminary information about arterial oxygenation. We demonstrate that the ToF data is independent of the intensity of the laser source, and hence demonstrate the potential of obtaining the hemodynamic oxygenation directly from temporal information which leads to minimizing the effect of intensity variations such as different skin tones which has a large variation among the population. Our prototype instrument includes three main parts: a laser source, a detector, and a Time-to-Digital Converter (TDC). We use both a commercially available pulsed laser source and a low-cost compact custom-developed laser source in order to compare the effect of optical power and temporal resolution in our measurements. The detector is a SPAD [7] which is a reverse biased PN junction sensitive to individual photons. With its picosecond temporal resolution, SPADs are an ideal photodetector for ToF measurements. The Time-to-Digital Converter (TDC) measures the time intervals between two events. In this prototype, the TDC is based on a Field Programmable Gate Array (FPGA) which enables greater flexibility and integration during the prototyping phase. To evaluate our result, we use the Vascular Occlusion Test (VOT) which induces a measurable change to the tissue oxygenation and total hemoglobin [9]. During the occlusion stage, the oxygenation levels fall slowly, and de-oxygenation levels rise. In this study, cuff occlusion is applied on the upper arm for the observation of changes in hemodynamic absorption, both in intensity and ToF data.

## 2. PROPOSED METHODOLOGY

We have developed a compact prototype instrument for the pulse oximetry experiments. Figure 1 shows a block diagram of this prototype which consist of the sensor unit for photodetection, the processing unit and the laser source. The sensor consists of a SPAD and the power management for biasing and buffering the SPAD output which is connected to the data processing unit. The processing unit includes the TDC, histogram generator and the required computational logic to transfer the collected data to the PC via a standard gigabit network. Additionally, the data processing unit provides a synchronization signal to the laser source.

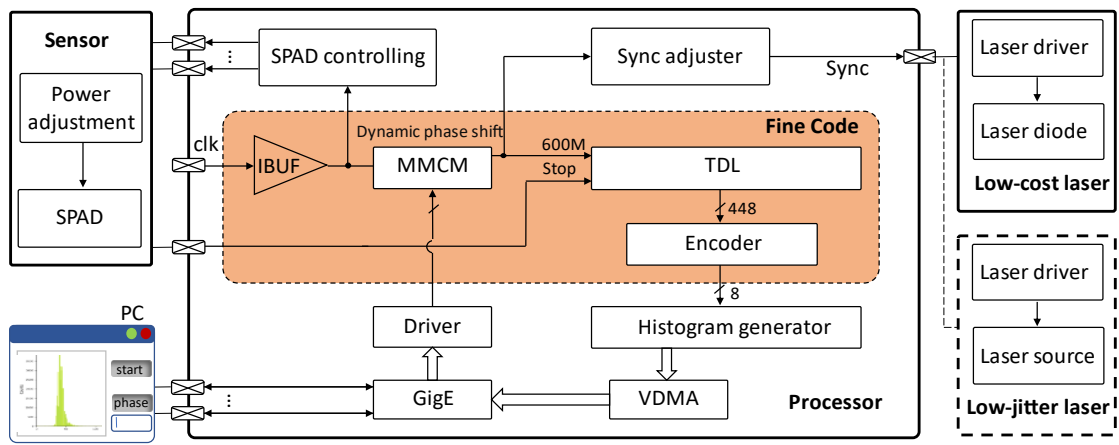

Figure 1. Block diagram of the prototype instrument

### 2.1 SPAD Sensor

The SPAD detector in this protype oximeter is an integrated passively quenched SPAD pixel [7] with a photo-sensitive area of 8μm diameter, a dark count rate of 1kcount/s, and a recovery time of approximately 20ns. The photon detection probability of this SPAD for the 773nm and 650nm laser used in this experiment is respectively 10% and 25%. We only use a single pixel to eliminate the effect of cross talk, signal and power integrity which may increase jitter of the detector.

### 2.2 Laser sources

In this study we use two laser source configurations. In the first configuration we use a Hamamatsu C10196 driver coupled with a Hamamatsu M10306 laser head of wavelength 773nm, a pulse duration minimum 56ps including temporal jitter. This configuration is a reference for validation of the prototype instrument. Subsequently, in the second configuration we use a custom developed low-cost and compact pulsed laser source based on the Texas Instruments LMG1025-Q1EVM laser driver evaluation board with a 650 nm Roithner LaserTechnik QL65D6SA laser diode. The evaluation board contains the following modules in-series: an electrical pulse shortener circuit, a LMG1025-Q1 transistor driver, and a EPC2212 GaN FET. Electrical pulses on the FET gate regulate the current through the laser diode.

The pulse shortener circuit in the evaluation board allows pulses of a longer width as input. It is made of an AND gate with one of its inputs connected directly to the input sync signal and the other connected to the same signal through a RC circuit delay generator. Intuitively, the width of the output pulse is the difference between the time of the input sync signal at logic '1' and the delay caused by the RC circuit. The pulse electrical shortening is caused by the RC circuit delaying the time of both AND gate inputs reaching the logic '1'. The width of the output pulse is then the time between the time instant of the AND gate outputting a logic '1' and the time instant when the input square wave drops to logic '0'. The input we use is a 4 MHz pulse waveform with a 61% duty cycle (153ns high time). The width of the output electrical pulse is approximately 1.3ns. The laser diode is operated at the gain-switched mode which means that the output optical pulse is significantly shorter than the input electrical pulse to the laser diode [10]. The optical pulse duration and total jitter of this low-cost laser source is measured as part of our experiments.

## 2.3 Time to digital converter

The TDC is implemented in a Xilinx Zynq UltraScale+ FPGA using the Tapped Delay Lines (TDLs) architecture with 45ps bins. The TDL consists of carry chain blocks and flip-flops. Propagation states are sampled when a corresponding photon event occurs. The TDL has a total of 448 delay taps and a measurement range of 1.66ns. The linearity of the TDL improves with the increasing bin width. In these experiments, the largest available bin width of 45ps was selected due to a better linearity which is required for accurate ToF measurements. The linearity is estimated by computing Differential Non-Linearity (DNL) and Integral Non-Linearity (INL) values with a code density test [11]:

$$DNL(i) = \frac{I(i) - I(Avg)}{I(Avg)} \quad (1)$$

$$INL(i) = \sum_{j=1}^{j=i} DNL(i) \quad (2)$$

Where $I(i)$ is the count number of individual bins which represents the relative bin width, while $I(Avg)$ is the average count number of the entire TDL.

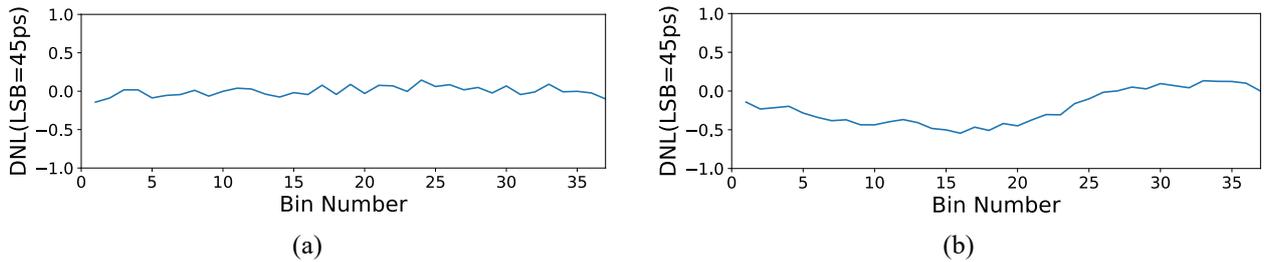

(a)　　　　　　　　　　　　　　　　(b)
Figure 2. Code density test results of the 45ps TDC with 37 bins. (a) DNL plot and (b) INL plot

Figure 2 shows the results from the code density test which demonstrates that DNL and INL are respectively [-0.15, 0.15] Least Significant Bit (LSB) and [-0.55, 0.15] LSB without missing codes and without bubbles [11]. A FPGA-based histogram generator is implemented with 37 bins of a 16-bit depth each. Frame rate control is within the histogram generator which indicates the start and stop of each histogram. A 600Mhz clock is used as the start signal of the TDC, and a single clock cycle is used as the maximum measurement range which is 1667ps. The 37 delay steps are arranged in one measurement cycle which means the average bin width is 1667ps/37=45ps.

## 2.4 Data calculation

We use equation (3) and (4) to calculate the mean time <t> of the temporal spread function (TPSF) based on a second order Mellin-Laplace transform [12]. The mean time calculation is sensitive to the position and shape of the TPSF. However, <t> remains unaffected by intensity fluctuations, since the shape of the TPSF does not depend on intensity.

The <t> for discrete data of the TSPF is calculated using:

$$W(i) = \sum_{i=0}^{n} I(i) \qquad (3)$$

$$<t> = \sum_{i=0}^{n} if(i) = \sum_{i=0}^{n} i * I(i) / W(i) \qquad (4)$$

where $I(i)$ is the individual bin counts and $W(i)$ is the sum of counts over all bins. $n$ is the total number of bins while $<t>$ is the mean flight time of the photons.

**2.5 Vascular Occlusion Test Protocol**

In order to evaluate the performance of our prototype in terms of measuring oxygenation levels in the blood, we use the Vascular Occlusion Test [8][9], which uses a manually pressurized cuff on the upper arm to reduce the blood flow to the lower arm and fingers. In these experiments we monitor the transmission of the laser light through the volunteer's index finger. A baseline is provided prior to the start of cuff pressurization. Once started, the cuff is pressurized to 160 mmHg and maintained for a few seconds. Then the cuff's pressure is released rapidly, so that the hemodynamics can return to the baseline value prior to the cuff pressurization.

Depending on the position of the arm and finger relative to the rest of the body, the hemodynamic may increase or decrease, as the total hemoglobin volume changes. In return, the absorption levels in the blood hemodynamics changes which leads to intensity fluctuations. These fluctuations are large enough to be observable for prototyping pulse oximetry instruments. In the ToF measurement, the change in the absorption is measured by the change in shape of the TPSF, and it is quantified by changes in the mean-time <t> calculations. For example, a higher total absorption leads to TPSF with a lower temporal spread, hence lower <t>. An advantage of calculating the mean-time measurement instead of intensity is that mean-time is independent of the intensity, which is the area under the TPSF.

In these experiments we used a single wavelength as it provides sufficient information for proof-of-concept ToF measurements. Further wavelengths are required in order to determine the oxygenation concentrations. However, the demonstrated measurement method remains identical for various wavelengths.

## 3. EXPERIMENTS AND RESULTS

We designed two separate setups for the measurement of photon intensity and mean-time <t>, one with the Hamamatsu laser and the other with the low-cost laser. Figure 3 shows the experiment's setup for the prototype instrument. For both Figure 3(a) and Figure 3(b), the laser source is located on the left side of the image and the SPAD detector on the right side of the image. In this experiment the SPAD sensor detects photons from the laser source which are transmitted through the tissue. The electronic board consisting of the SPAD sensor and the data processing unit is approximately 10cm×8cm×5cm, and the low-cost laser source module has dimensions of 4cm×5cm×0.5cm. Each laser has its own power setting, and it is synchronized to the TDC with a 4MHz pulse.

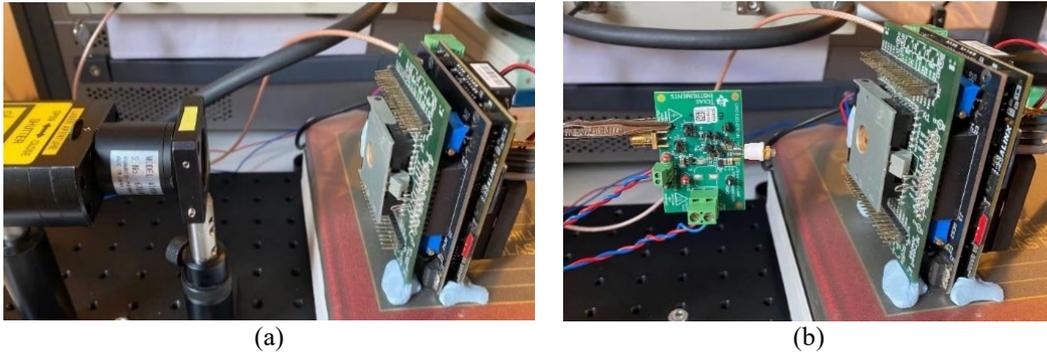

(a)                          (b)

Figure 3. Experiment's setup for the measurement for the protype instrument showing the electronic board with (a) Hamamatsu laser (b) low-cost laser.

## 3.1 Measurement with Hamamatsu laser source

The measurement stability of the prototype instrument is tested with the Hamamatsu laser source. Figure 4 shows the results of this test which has 45k frames of data are recorded and transmitted to the PC at a rate of 50 frame/s. The intensity is calculated from the area under the TPSF for each frame. The mean-time <t> is calculated based on section 2.4. During the experiment, the average optical power is approximately 0.5μW, and all the components are kept in a static position. The relative change in intensity and mean-time are respectively 41% and 3% during the 15 min stability test. The experiment shows the drift in mean time measurement has a better robustness than the intensity measurement under the same conditions.

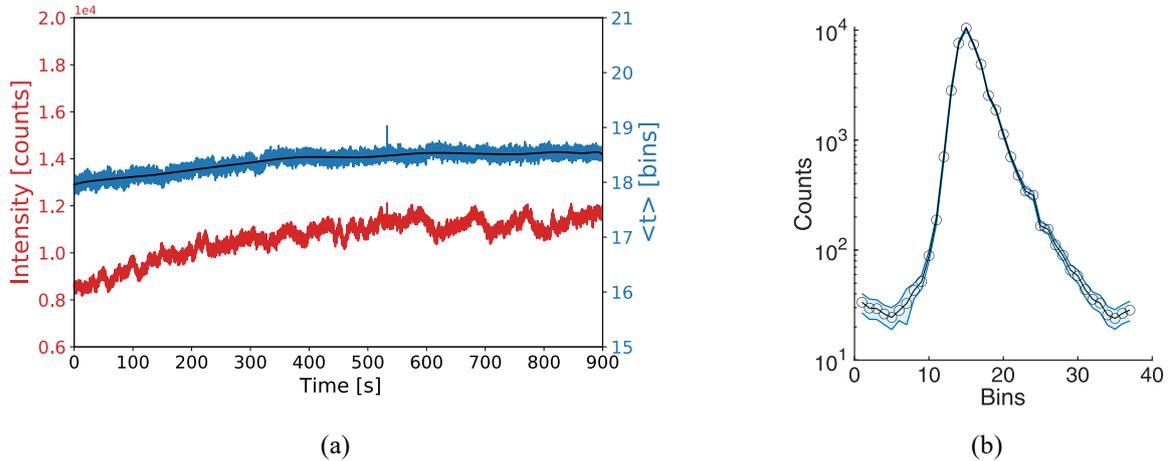

Figure 4. System stability test for the Hamamatsu laser. 45k frames of data are collected at 50 frame/s. (a) drift in raw intensity and mean-time data (b) average IRF showing the standard deviation as the blue shadow area. The FWHM of the IRF is approximately 3.6bins (=165ps)

In order to evaluate the independence of the mean-time <t> measurement from the intensity fluctuations, three different neutral density filters are sequentially introduced to the light path between the laser source and the SPAD sensor. The three filters are Thorlabs NE10A, NE20A-B, and NENIR40A-C. Figure 5 shows the results of this experiment which has a total change in the intensity of 138% whilst the ratio of the mean-time change is 1% indicating the independence of ToF to intensity of the laser source. The average optical power measured on the SPAD without filters is approximately 20μW.

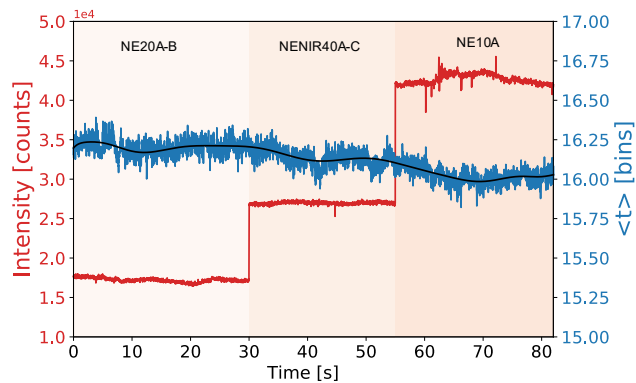

Figure 5. Intensity variations using different optical densities filters with Hamamatsu laser source showing independence of mean-time measurement from intensity measurements. Intensity data is displayed in red and <t> is displayed in blue color. The black line is the fitting for visualization. Frame rate is 50 frames/s, and bin is 45ps.

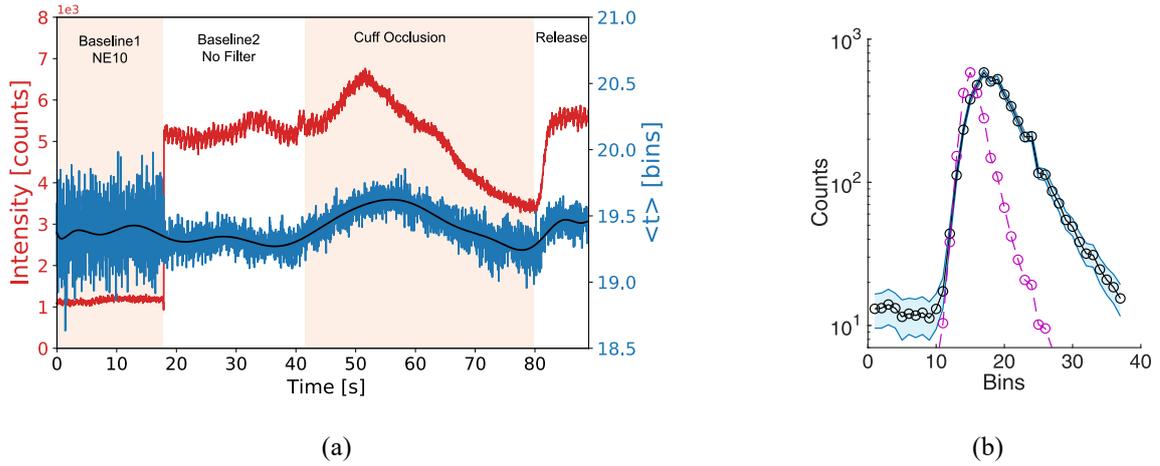

Figure 6. The results from hemodynamic changes during multiple stages of the Vascular Occlusion Test (VOT) showing with the Hamamatsu laser (a) intensity in red and mean-time <t> in blue. Baseline1 has Thorlabs NE10A, and in Basline2 the filter is removed, the black line is polynomial data-fitting for visualization. Data rate is 50 frames/s, and bin is 45ps (b) average TPSF during the baseline with no filter is displayed in black, standard deviation is displayed in shadowed blue region, and average IRF is displayed in dashed purple.

We perform the Vascular Occlusion Test (VOT) with Hamamatsu laser evaluating the functionality of our prototype. Figure 6 shows a 100s test which contains baseline1, baseline2, cuff occlusion, and release. In order to ensure the independence of the <t> from the intensity data, the Thorlabs NE10A filter was added at the beginning of the test during the baseline1 section. This filter was removed for the reminder of the baseline. Whilst there is clear change in the intensity data, there is no observable changes in the trend of <t>. However, it is noticeable that the noise of the measurement has increased during the baseline1 section. This is due to a higher shot noise because of receiving less photons as the intensity of the laser source is reduced by the ND filter. The average optical power measured on the finger without filters is approximately 6μW.

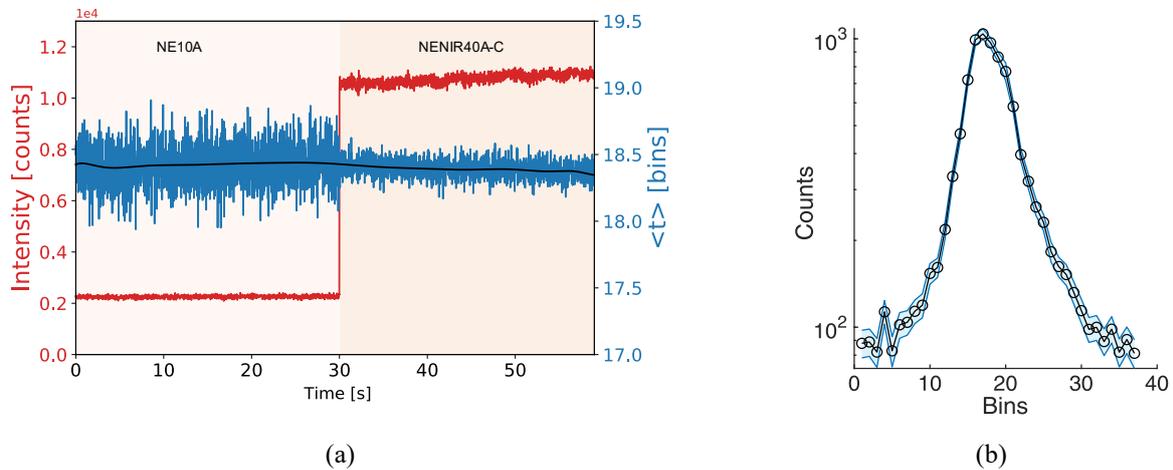

Figure 7: Intensity variations with different neutral densities filters Thorlabs NE10A and NENIR40A-C with the low-cost laser source showing independence of mean-time measurement from intensity measurements. (a) Intensity data is displayed in red and <t> is displayed in blue color. The black trend is fitting for visualization. Data rate is 50 frames/s, and bin=45ps (b) average IRF when NENIR40A-C ND filter is used is displayed is black and the standard deviation is displayed as the shadowed region in blue. The FWHM of the average IRF is approximately 7bins (=315ps).

## 3.2 Measurement with low-cost laser source

Since we have demonstrated the stability of our prototype instrument with the Hamamatsu laser in the previous section, we provide a stability test with the custom developed low-cost laser source. Figure 7 shows the results of the stability test for the low-cost laser. Two Thorlabs ND filters are used to change the intensity during this stability test. The total change in the mean-time <t> data is 1% whilst the intensity data changes by 360% during this test. Similar to Figure 6, the <t> has a higher noise due to lower number of photons. Despite a similar optical power with the Hamamatsu laser, the low-cost laser has a lower dynamic range due to a higher baseline as the laser does not fully switch OFF to achieve a reasonably short optical pulse. This dynamic range difference is observable in Figure 4b and Figure 7b.

Therefore, we use our low-cost prototype laser source for the Vascular Occlusion Test. The average optical power measured on the finger is approximately 4µW. Since this optical power is of the low-cost laser diode is more diverging than the Hamamatsu laser, the low-cost laser source is moved closer to the volunteer's finger to get a better light coupling. Figure 8 shows the results from the Vascular Occlusion Test with the low-cost laser. As expected, the hemodynamic changes due to cuff occlusion are clearly observable on the intensity data, however, the noise in the mean-time <t> measurement is larger than the signal changes induced by the hemodynamic changes. The higher noise in the <t> measurement is due to lower number incident photons relative to the previous tests with Hamamatsu laser. Additionally, the drift observed in the <t> is potentially due to the drift in laser driver circuits such as temperature of the laser diode. It is noticeable that in both Figure 6 and Figure 8, that the amplitude the heartbeat signal is reduced during the cuff occlusion.

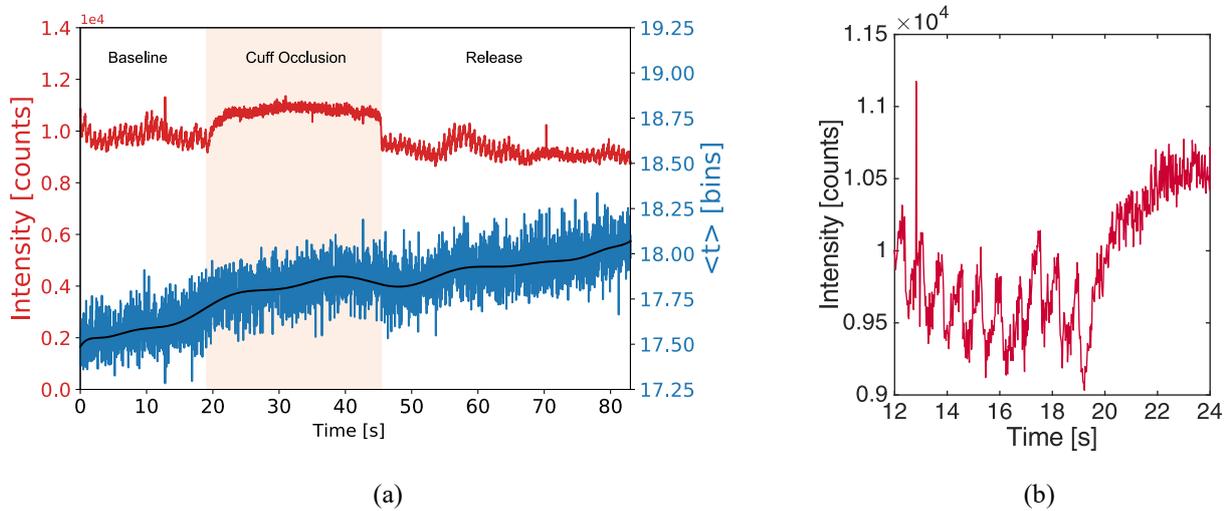

(a)           (b)

Figure 8: The results from hemodynamic changes during multiple stages of the Vascular Occlusion Test (VOT) showing with the low-cost laser (a) intensity in red and mean-time <t> in blue. The black line is polynomial data-fitting for visualization. Data rate is 50 frames/s, and bin is 45ps (b) The intensity data showing the heartbeat (=80bpm) signal at the beginning of the cuff occlusion.

## 4. DISSUCSION

The results between the Hamamatsu laser and low-cost laser show a significant decrease in the Signal-to-Noise (SNR) ratio of the mean-time <t> measurement. Although the detected photon count between the two lasers remains similar, the background photon count is approximately 10 times higher than Hamamatsu laser. This is potentially due to the biasing of the laser diode within the laser driver circuit. There is a trade-off between laser power, and pulse width, and subsequently achieving the appropriate biasing condition requires further investigation. As a result, a large number of detected photons, and lower background photons are required to improve the SNR. This could be achieved using a SPAD with a higher Photon Detection Probability (PDP), and lower Dark Count Rate (DCR). However, due to limited recovery time of the SPAD, a higher number of photons may cause a pile-up and lead to false TPSFs. Both PDP and DCR are part of the avalanche diode characteristics which depend on the semiconductor manufacturing processes. Another method to increase the number of detected photons is increase the photon detection area which could be achieved by creating an array of

SPAD pixels [11]. Despite a larger photon detection area, other characteristics such as low crosstalk, temporal jitter, and temporal drift remain a challenge in large arrays of SPADs.

Another strategy is to increase the number of emitted photons by using a higher laser power, and increased repetition rate. This could be achieved by providing a larger electrical pulse to the laser diode. However, due to the gain-switching nature the higher optical power may lead to a larger optical pulse duration, widening the IRF [10]. An alternative method to increase the number of emitted photons is to increase the repetition rate of the laser source which depends on the electronic circuit in the laser driver. In this case, the TDC is also required to record events at these higher repetition rates, and the SPAD's recovery time should be faster than the repetition rate to avoid pile-up.

Relative to an intensity-based pulse oximeter, a ToF pulse oximeter includes more complex components which require further improvement with the existing technology. Once improved, a ToF based pulse oximeters potentially provides a better robustness to variations such as intensity of the laser source and changes in the skin tone among the population. Furthermore, it is possible to extract the absolute oxygenation level directly from the measured TPSF.

## 5. CONCLUSION

We have performed a series of experiments to demonstrate the changes in measurements of hemodynamics during Vascular Occlusion Test. Additionally, we have demonstrated the independence of the mean-time measurement from intensity fluctuations in our prototype instrument. This has the potential to lead to pulse oximeter instruments which are based on ToF measurements, rather than the intensity based, which is susceptible to variations in the incident light, hence providing more robust oximetry measurements. Furthermore, the absolute values of scattering and absorption could be extracted from the TPSF data using deconvolution techniques, especially when a shorter laser pulse duration is used. Given a sufficiently high SNR, it may be possible to measure the absolute variations in the oxygenation levels using such low-cost and compact instruments which can be embedded into everyday devices such as wearable devices and smartphones.

## ACKNOWLEDGMENTS

The authors would like to thank QuantIC project (https://quantic.ac.uk/) for funding this work (EPSRC grant number EP/T00097X/1). This work was partly supported by National Natural Science Foundation of China (NSFC) under Grant 11603080. We especially thank Alistair Gorman for helpful discussions, Hanning Mai for technical suggestions, Yangchun Li and Shanny Lee for their assistance in the experimental setup.